%% file: paper.tex
\documentclass[prl,aps,twocolumn,groupedaddress,superscriptaddress,preprintnumbers,nobibnotes]{revtex4-2}
\include{ethans_preamble.tex}

\usepackage[caption=false]{subfig}
\usepackage{overpic}
\usepackage{tikz}
\newcommand{\labeledfig}[2]{%
	\begin{tikzpicture}[inner sep=0pt, outer sep=0pt,
		baseline=(current bounding box.north)]
		\node[anchor=north west] (image) at (0,0) {\includegraphics[width=\linewidth]{#1}};
		\node[anchor=north west, inner sep=2pt] at (image.north west) {{\bf #2}};
	\end{tikzpicture}
}

\usepackage{enumitem}

\usepackage{amsthm}
\usepackage{physics}

\usepackage{bm}

\DeclarePairedDelimiterX{\infdivx}[2]{(}{)}{%
	#1\;\delimsize\|\;#2%
}

\newtheorem*{theorem*}{Theorem}

\usepackage{accents}

\newcommand{\oEE}{\mathop{\mathbb{E}}}

\newcommand{\tmem}{{t_{\sf mem}}}
\newcommand{\ler}{{\xi_{\sf er}}} 
\newcommand{\dr}{{\sfD \sfR}}
\makeatletter
\newcommand\blfootnote[1]{%
	\begingroup
	\def\@thefnmark{}%
	\@footnotetext{#1}%
	\endgroup
}
\makeatother
\begin{document}

	\title{Long-lived memory in sliding spin chains}
	\author{Charles Stahl} 
	\affiliation{Department of Physics, Stanford University, Stanford, CA 94305, USA}
	\author{Ethan Lake}
	\email{elake@berkeley.edu}
	\affiliation{Department of Physics, University of California Berkeley, Berkeley, CA 94720, USA}
	\begin{abstract}
		We study a system of two ferromagnetic one-dimensional Ising chains coupled to a thermal bath, which are driven out of equilibrium by being moved past one another at a constant speed. We show that even at modest speeds, magnetic friction between the two chains significantly increases the ability of the system to order. In particular, at inverse temperature $\beta$, Ising coupling $J$, and sliding speed $v$, the dynamics retains memory of its initial magnetization for a time that increases from $\exp(O(\beta J))$ at $v = 0$ to $\exp(O((\b J)^2v))$ when $v>v_c$, where $v_c$ is a small constant. Magnetic friction thus provides a simple mechanism for parametrically slowing down thermalization in a one-dimensional magnet. 
	\end{abstract}
	
	\maketitle
	
	Ordered phases of matter retain memory of their initial conditions: prepare a ferromagnet with a particular magnetization, and at low enough temperatures, its magnetization will remember its initial value for thermodynamically long times. Historically, order has been understood through the lens of spontaneous symmetry breaking. However, approaching order from the perspective of memory is better suited for understanding {\it non}equilibrium phases of matter, some of which are capable of retaining memory in the absence of {\it any} symmetry~\cite{toom1974nonergodic,toom1980stable,gacs1978one,gacs2001reliable,cirel2006reliable}. Given the outstanding difficulties in classifying nonequilibrium phases of matter, and the immense technological importance of magnetic memories~\cite{tang2010magnetic}, understanding how to construct physical systems that are good at remembering information is a matter of both fundamental and applied importance. While it is an old subject~\cite{toom1974nonergodic,toom1980stable,gacs1978one,gacs2001reliable,cirel2006reliable}, it has also seen several recent advances~\cite{mordvintsev2020growing,pajouheshgar2026exploring,lake2025squeezing,stahl2026slow,stahl2026brownian}. 
	
	Memory in one dimension, where equilibrium systems cannot order, is both especially interesting, and especially poorly understood. G\'acs (in)famously constructed an example of a 1d cellular automaton that retains memory for thermodynamically long times in the absence of any symmetry~\cite{gacs2001reliable}, but his construction---based on an infinite hierarchy of self-simulating Turing machines---is so complicated that it has yet to be independently understood. Attempts to construct simpler models have so far come short~\cite{gacs1978one,toom1995cellular,park1997ergodicity}. An outstanding question thus remains: are there {\it simple} ways for 1d systems to retain memory? Or, failing this, is there at least a simple model which gets close? 
	
	This work addresses the last question by constructing a 1d spin system that, when driven out of equilibrium in a simple way, remembers its initial conditions for a very long time, even in the presence of noise. While this time is finite in the thermodynamic limit, it is nevertheless significantly longer than any 1d equilibrium system can manage, in a sense made precise below. 
	
	The construction in our paper is a spin system that retains a single bit of memory in the sign of the average magnetization $m$. In this setting, we define the {\it memory time} $\tmem$ as the expected time for a maximally-magnetized state to lose memory of its initial magnetization: 
	\be \tmem = \oEE \min_{s = \pm 1}  \{ t \, : \,  |m(t)| <1/2 \, , \, m(0) = s\} \ee 
	where the constant $1/2$ is chosen rather arbitrarily. $\tmem$ can be viewed as a proxy for the mixing time, for which it sets a lower bound~\footnote{Up to a factor of the system size, if one uses the usual convention for the mixing time in the mathematics literature.}.
	
	To orient ourselves about the behavior of $\tmem$, consider Glauber dynamics run at temperature $T$ for the 1d Ising chain with coupling $J$. Once a minority domain (of any size) is formed, its endpoints will execute unbiased random walks. Since the energy cost to nucleate a minority domain is $\Delta E = 4J$, it follows that $\tmem \sim e^{4\b J}$~\cite{levin2026markov}. More generally, for any 1d spin system undergoing Glauber dynamics with respect to a local Hamiltonian $H$, a standard Arrhenius-type argument shows that $\tmem \lesssim  e^{O( \b J)}$, where $J$ is the largest local energy scale in $H$~\cite{bovier2016metastability,temme2017thermalization}. To get a longer memory time than this, we must go out of equilibrium. 
	
	\begin{figure*}[t]
		\centering
		\begin{minipage}[t]{0.32\textwidth}
			\labeledfig{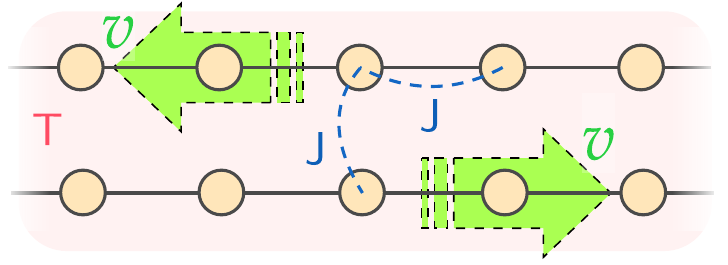}{a)}\\[1ex]
			\labeledfig{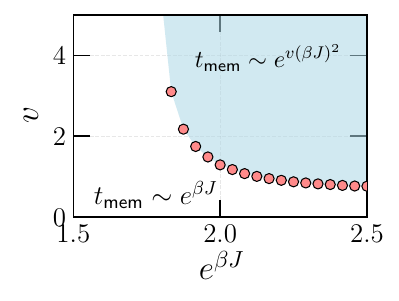}{b)}
		\end{minipage}\hfill
		\begin{minipage}[t]{0.32\textwidth}
			\labeledfig{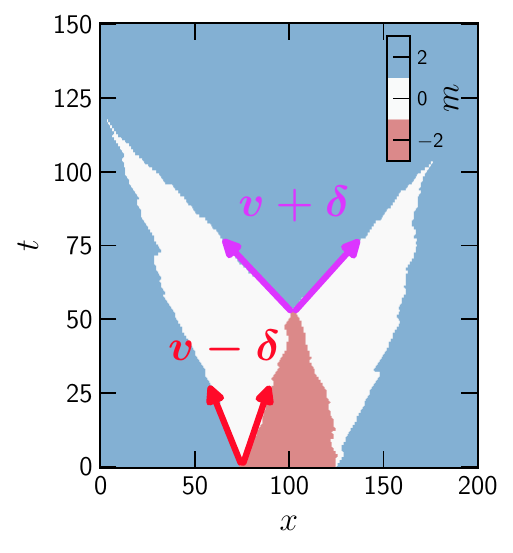}{c)}
		\end{minipage}\hfill
		\begin{minipage}[t]{0.32\textwidth}
			\labeledfig{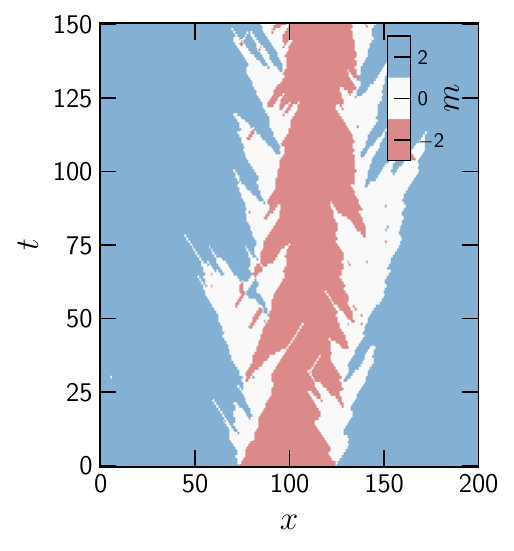}{d)}
		\end{minipage}
		\caption{\label{fig:overview} {\bf a)} Schematic of the dynamics. Two ferromagnetically coupled ferromagnetic 1d Ising chains with coupling $J$ slide past one another, each at velocity $v$, and both couple to a bath at temperature $T$. {\bf b)} Scaling of the memory time in the $e^{\b J}$-$v$ plane. In the white region, the memory time scales like $e^{O(\b J)}$, just as in the equilibrium 1d Ising model. In the blue region, the memory time is enhanced to $e^{O(v (\b J)^2 )}$. The pink points mark the crossover between the two, and are drawn by fixing $e^{\b J}$, and then finding the smallest value of $v$ such that $\tmem(v) >  10 \tmem(0)$ (with the constant $10$ chosen rather arbitrarily). {\bf c)} A spacetime history of the magnetization, with $\b = \infty, v = 1$. The indicated purple and red arrows denote interfaces that move with expected speeds $v + \d$ and $v-\d$, respectively. {\bf d)} A history with $\b = 2, v = 1$, with the minority domain in the initial state having size larger than $\ler$, the length scale below which minority domains are ballistically eroded. }
	\end{figure*}
	
	\paragraph*{Sliding Ising models:} In this paper, we show that subjecting a ferromagnetic 1d Ising model on a two-leg ladder to a very simple type of nonequilibrium drive can significantly increase $\tmem$. We drive the system out of equilibrium by sliding the two legs of the ladder past one another at velocities $\pm v$, as illustrated in Fig.~\ref{fig:overview}~{\bf a}. More precisely, letting $s^{(\sft / \sfb)}_i \in \{ \pm 1\}$ be the $i$th spin on the top / bottom leg of the ladder, we consider a system governed by the time-dependent Hamiltonian~\footnote{The floor and ceiling functions in this expression are not important, and can be replaced by a smoother kernel like $\sum_i \sum_j f(i-j+vt) s_i^{(\sft)} s_j^{(\sfb)}$ with e.g. $f(x) \sim e^{-x^2/{2\s^2}}$ without qualitatively changing the physics. } 
	\be \label{ham} H(t) = - J \sum_{i;a\in\{\sft,\sfb\}} s^{(a)}_i s^{(a)}_{i+1} - J \sum_i s^{(\sft)}_{\lfloor i-vt \rfloor} s^{(\sfb)}_{\lceil i+vt\rceil}  .\ee 
	We use updates where a configuration $C$ is replaced by a configuration $C'$ according to a rate $w(C\ra C')$ set by Glauber dynamics at temperature $T$:
	\be w(C \ra C') = \g (1 + e^{\b (E_{C'}(t) - E_C(t))})\inv, \ee 
	where $E_C(t)$ is the energy of configuration $C$ under $H(t)$, and $C'$ differs from $C$ by a single spin flip. In the following, we will fix the units of time so that $\g=1$~\footnote{In typical magnetic systems, the time for a local moment to relax to its Weiss field is around $10^{-8}$s~\cite{angst2012strongly}. Thus in our units, if we use a lattice spacing of $10^{-10}$m, a velocity of $v=1$ corresponds to a dimensionful speed of around $1 \,{\rm cm}\, {\rm s}\inv$.}. 		
	Models of ``magnetic friction'' like this have been studied before~\cite{kadau2008magnetic,hucht2009nonequilibrium,hilhorst2011two}, but past work has largely focused on the $v \ra\infty$ limit, and has not investigated how sliding impacts memory times~\footnote{It was however observed in~\cite{kadau2008magnetic} that sliding two 3d Ising models along a 2d interface can increase the magnetization near the interface. The present paper investigates the dynamics of a similar phenomenon in 1d.}.

	When $v=0$, the dynamics is simply that of a static two-leg Ising chain, and $\tmem \sim e^{O( \b J)}$. The main result of this work is to show that $\tmem$ is parametrically increased when $v$ is greater than a constant threshold. In particular, we give analytic arguments---and verify in numerics---that as long as $v > v_c$,  
	\be \label{main_result} \tmem \sim  e^{4(\b J)^2 v   + O(\b J)},\ee
	where the $\sim$ hides constant factors (see Fig.~\ref{fig:overview}~{\bf b} for a schematic phase diagram). We will argue below that $v_c \leq 1$ at large enough $\b J$. 
	
	While $\tmem$ is finite in the thermodynamic limit for all $T > 0$, it rapidly gets astronomically large as $T$ is decreased. For example, when $T = 0.6J$ and $v=2.5$, we find numerically that $\tmem \sim 10^{28}$, around 21 orders of magnitude larger than when $v=0$ at the same temperature~\footnote{In dimensionful units, typical parameters for magnetic solids would give a lifetime of $\approx 2.5\times 10^{11}$yr at $T \sim 150$K and a sliding velocity of $v \approx 5$cm ${\rm s}^{-1}$~\cite{angst2012strongly}.} (see below for more details). This enhanced lifetime does not depend on the $\zt$ spin-flip symmetry in~\eqref{ham}, and breaking it with a small magnetic field $h$ simply results in the replacement $J \mt J-|h|$.
	The remainder of this paper is dedicated to explaining this result.

	\paragraph*{The shearing mechanism: } In the equilibrium 1d Ising model, once minority domains are created, there is nothing that compels them to shrink. Instead, they expand diffusively, and quickly erase memory of the initial magnetization. 
	To understand why sliding changes this, consider what happens to minority domains under the sliding dynamics. Due to the interchain coupling, a minority domain on only a {\it single} chain is eroded ballistically from its endpoints, which violate two ferromagnetic bonds. We will write $\d$ for the velocity at which the endpoints of a minority domain on a single chain contract; at small $T$ the Glauber transition rates give $\d \approx 1 / (1 + e^{-2 \b J})$. Note that single-chain minority domains are eroded ballistically even in equilibrium, when $v=0$.
	
	A minority domain that spans {\it both} chains is more problematic. When $v=0$, its endpoints diffuse, and nothing in the dynamics compels it to shrink. When $v>0$ however, the sliding dynamics attempts to shear the domain, ripping it apart into two spatially-separated single-chain minority domains. The ferromagnetic interchain coupling attempts to arrest this shearing, as it favors configurations where domains on each chain ``stick'' together. But if $v >\d$ and $T$ is low enough, it is easy to check that the domains are sheared apart faster than they stick together, and once sheared apart, they are ballistically shrunk by the mechanism discussed above. This process, viz. the ripping apart of a two-chain minority domain and the subsequent ballistic erosion of its two constituent single-chain halves, is shown in Fig.~\ref{fig:overview}~{\bf c}, and is what leads to the aforementioned increase in $\tmem$. 
	
	In passing, we remark that this shearing mechanism produces dynamics that are qualitatively similar to Toom's two-line-voting automaton~\cite{toom1995cellular,park1997ergodicity} and the GKL automaton~\cite{gacs1978one}, two noisy cellular automata which were originally proposed as candidates for one-dimensional memories. This was shown not to be the case in~\cite{park1997ergodicity}, which proved that dynamics subjected to $\zt $-symmetry-breaking noise of strength $\ep$ has $\tmem = e^{\ct(\log^2(\ep))}$ (it was conjectured from numerics in~\cite{de1992gacs} that $\tmem = e^{\ct(1/\ep)}$ for symmetric noise, but our analysis will show that in fact the $e^{\log^2(\ep)}$ scaling holds in this case as well; see~\cite{code} for details).
	
	How effective is the shearing mechanism? First consider what happens when $v \ra \infty$: here, domains are sheared apart instantly, and a spin on one chain couples to spins on the other chain in a random order. 
	The inter-chain interactions effectively become all-to-all, and the model enters a mean field limit, where it is easy to show the existence of a phase with long range order (see e.g.~\cite{kadau2008magnetic,hucht2009nonequilibrium}). In this regime, a system of size $L$ has genuine long-range order, with $\lim_{L \ra\infty}\tmem = \infty$ at low enough $T$ (and in fact we only need $v = \O(\log L)$ for this to be true). 
	
	At finite $v>\d$, we will argue that $\tmem$ scales as in~\eqref{main_result} at low enough $T$ (at small $T$, $\d\leq1$ thus sets an upper bound on the parameter $v_c$ that controls  when \eqref{main_result} holds). This argument will proceed in the following stages. First, we will show that two-chain minority domains (referred to as just ``minority domains'' in the following) are sheared apart with high probability as long as their size $l$ satisfies $l < \ler$, where the {\it erosion length} $\ler$ is a length scale we will show behaves as $\ler \sim v e^{\b J}$. We will also show that minority domains of size $l > \ler$ are {\it not} ballistically eroded, and that their endpoints instead undergo diffusive motion, as is the case for domains in the equilibrium 1d Ising model (see Fig.~\ref{fig:overview}~{\bf d} for an example). $\tmem$ is thus set by the time it takes for the dynamics to spontaneously nucleate a minority domain of size $\ler$. We will show~\eqref{main_result} by showing that, once a minority domain of a small constant size is created, the probability that it expands to size $\ler$ scales as $e^{O(\b J \ln \ler )}$. Finally, we will then check in numerics that both of the steps in this argument are correct. 
	
	\begin{figure}
		\centering 
		\includegraphics[width=.45\tw]{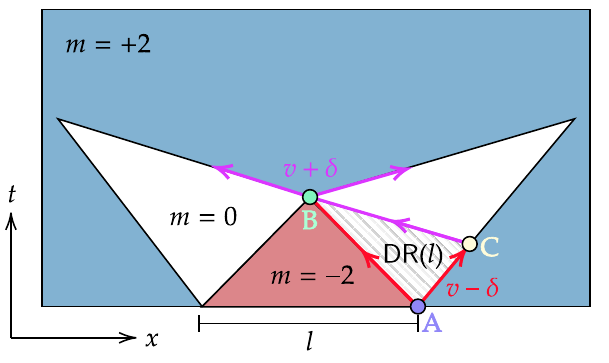}
		\caption{\label{fig:proof} How a minority domain of size $l$ is eroded under ideal sliding dynamics with $v >1$ in the limit $\b J \ra \infty$. The white regions have zero magnetization, and are metastable. The red arrows indicate interfaces that move with expected speed $v-\d$, and the purple arrows those with speed $v+\d$. The danger region $\dr(l)$ indicated by the shaded triangle is the spacetime region where spin flips in the metastable zero-magnetization area can propagate to prevent the erosion of the $m=-2$ (red) part of the initial minority domain. } 
	\end{figure}
	
	\paragraph*{The erosion length: } 
	We begin by establishing the scaling of the erosion length $\ler$. Fig.~\ref{fig:proof} shows a schematic of how minority domains are eroded in the $\b J \ra \infty$ limit. The white triangles are spacetime regions with $m=0$ where the initial domain has been sheared apart, with minority spins present on only one of the chains. If a majority spin flips to match a minority spin in this region---which happens with probability $\sim e^{-2\b J}$---it can rapidly nucleate a new domain with minority spins on both chains. These processes are what prevent the erosion of minority domains of sufficiently large size $l$, as when $l$ is large enough, a majority spin flip occurring somewhere in this spacetime region is a likely event. When these events are likely, they result in the minority domain endpoints executing diffusive random walks, as exhibited in Fig.~\ref{fig:overview}~{\bf d}.
	
	Let us more carefully consider the spacetime region where events that flip majority spins can lead to the minority domain directly increasing in size. Consultation of Fig.~\ref{fig:proof} shows that this happens in the hatched spacetime region marked as ${\sf DR}(l)$, which we will refer to as a {\it danger region}. $\ler$ can be estimated as the size $l$ where having a majority spin flip somewhere in $\dr(l)$ becomes likely: 
	\be \label{ler_cond} e^{-2 \b J}|\dr(\ler)| \sim 1.\ee 
	$|\dr(l)|$ can be determined from simple geometric considerations as $|\dr(l)|\approx l^2 \d / (4v(v-\d))$~\cite{code}. From~\eqref{ler_cond}, we then have 
	\be \label{ler_scaling} \ler = c v e^{\b J}, \ee 
	where $c$ is a constant independent of $v$ to leading order in $\d/v$. 
	
	\begin{figure*}[t]
		\centering
		\begin{minipage}[t]{0.24\textwidth}
			\labeledfig{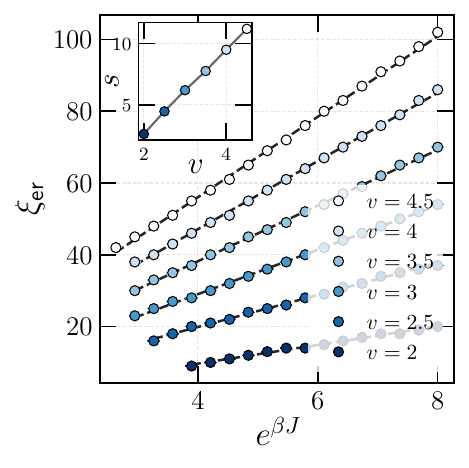}{a)}
		\end{minipage}\hfill
		\begin{minipage}[t]{0.24\textwidth}
			\labeledfig{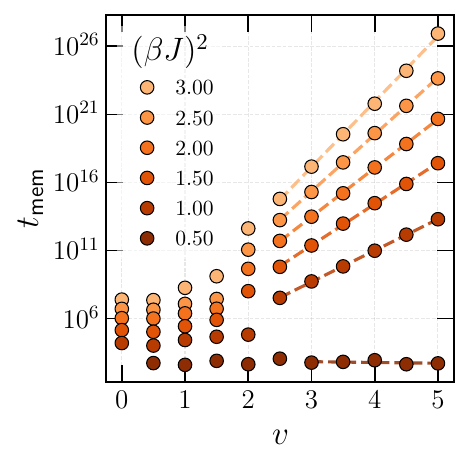}{b)}
		\end{minipage}
		\begin{minipage}[t]{0.24\textwidth}
			\labeledfig{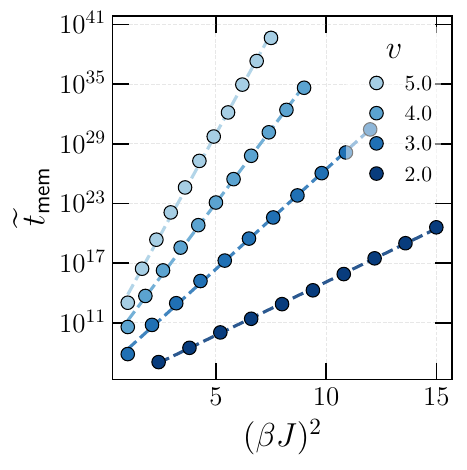}{c)}
		\end{minipage}\hfill
		\begin{minipage}[t]{0.24\textwidth}
			\labeledfig{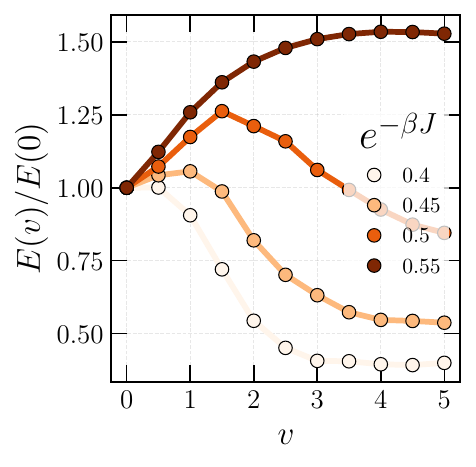}{d)}
		\end{minipage}
		\caption{\label{fig:quantfig} 
			{\bf a)} The erosion length $\ler$ against $e^{\b J}$ for different values of $v$. The inset shows the slopes $s$ of linear fits of $\ler$ to $e^{\b J}$ (drawn as dashed black lines), confirming that $\ler \sim v e^{\b J}$. 
			{\bf b)} $\tmem$ computed with forward flux sampling as a function of $v$, confirming that $\tmem$ scales exponentially with $v$ when $v > 1$ and $\b J$ is large enough. Dashed lines are linear fits of $\ln \tmem$ to $v$ in the range $v > 1$. 
			{\bf c)} Nucleationless memory time $\wt  t_{\sf mem}$ (see the main text) as a function of $(\b J)^2$, indicating that $\tmem$ indeed scales exponentially with $(\b J)^2$ when $v>1$ and $\b J$ is large enough. Dashed lines are linear fits of $\ln \wt t_{\sf mem}$ to $(\b J)^2$. 
			{\bf d)} Average energy $E(v)$ in the steady state of the sliding dynamics at speed $v$, normalized to the value $E(0)$ of the energy at $v=0$. At large enough $\b J$ and $v$, the energy {\it decreases} with $v$. In all plots, standard error bars are smaller than the data points. 
		}
	\end{figure*}
	
	To determine $\tmem$, we thus need to understand the most likely process by which fluctuations produce minority droplets of size $\ler$. If such droplets had to be created ``all at once'' by rare fluctuations, or had to be built up continuously from size $1$ to size $\ler$, we would expect $\tmem \sim (e^{O(\b J)})^\ler$. This estimate is too naive, however, as there is a more efficient way of building up minority domains. 
	
	To understand this mechanism, consult again the schematic in Fig.~\ref{fig:proof}. If a majority spin flips in $\dr$, the largest amount by which it can increase the size of the minority domain is $x_C - x_A$, where $x_{A/C}$ are the spatial coordinates of the points $\bfA,\bfC$ in the figure. Basic geometric considerations show~\cite{code} that this distance equals $l\d /2v$. A spin flip event with likelihood $\sim e^{-2\b J}$ can therefore increase the size of the minority domain by at most $l \ra l(1+\d /2v)$. Since the domain size increases multiplicatively, the number $n_\ler$ of such spin flips needed to grow the domain to size $\ler$ is $n_\ler = \ln(\ler) / \ln(1+\d/2v)$. We therefore expect 
	\be \tmem \gtrsim b(e^{2\b J})^{\frac{\ln (\ler)}{\ln(1+\d/2v)}} \sim be^{2(\b J)^2 \frac{1}{ \ln(1+\d /2v)}},\ee 
	for some constant $b>0$ (dropping the $\ln(v)$ part of $\ln(\ler)$). We have written $\gtrsim$ here because this argument ignores the time needed to nucleate a small minority domain that can then be expanded by the mechanism above, a time which scales as $e^{O(\b J)}$. Adding this in and expanding in $\d /v \ll 1$ then gives the main result~\eqref{main_result}~\footnote{It is interesting to note that this line of reasoning, specifically the part where we argue that $n_\ler \sim \ln(\ler)$, is similar to the approach used to show that the mixing time of the East model scales as $e^{O(\beta^2)}$~\cite{garrahan2011kinetically}.}. 
	
	The above argument has taken place in the presence of a $\zt$ symmetry, brought about by the lack of a symmetry-breaking magnetic field. The sliding-induced enhancement of $\tmem$ is however not dependent on this symmetry, and repeating the arguments above in the presence of a magnetic field simply results in the replacement $J \ra J - |h|$ in~\eqref{main_result}.

	\paragraph*{Numerics: }
	We now confirm these predictions in numerics, sticking for concreteness to the case with no magnetic field. We start with our prediction for $\ler$, which we estimate numerically as follows. First, we prepare an initial state containing a minority domain of length $l$ in a system of size $L = 5 v l$. We then evolve for a time $l$, and define the {\it erosion probability} $\sfP_{\sf er}(l)$ as the probability that the total number of minority spins has decreased to less than $l$ by this time. We then estimate $\ler$ as the largest value of $l$ where erosion happens with a probability greater than an arbitrary fixed constant, which we choose to be $3/4$ for concreteness: 
	\be \label{numeric_ler} \ler =  \max \{ l \, : \, \sfP_{\sf er}(l) > 3/4 \}.\ee
	Fig.~\ref{fig:quantfig}~{\bf a} shows the result of measuring~\eqref{numeric_ler} in numerics, and confirms the predicted scaling~\eqref{ler_scaling} at large enough $\b J, v$. 
	
	We now confirm that $\tmem$ indeed scales as in~\eqref{main_result}. Doing so is infeasible with direct Monte Carlo updates, since the estimated values of $\tmem$ quickly become extremely large. Fortunately, forward flux sampling (FFS)~\cite{allen2009forward}---a technique that selectively isolates updates that move the magnetization out of its metastable minimum---allows us to get reliable estimates for $\tmem$, even in this setting. For details about FFS and the full codebase used to perform the numerics, see~\cite{code}. 
	
	Fig.~\ref{fig:quantfig}~{\bf b} shows the FFS estimate of $\tmem$ plotted against $v$ for different values of $\b J$. As expected, we observe $\tmem$ to grow exponentially in $v$ when $v>1$ (and perhaps at still smaller values of $v$). 
	Confirming the scaling of $\tmem$ with $\b J$ is slightly more subtle. As discussed above, the seeding event where a small minority domain is nucleated contributes a factor of $e^{c\b J}$ to $\tmem$. Creating a minority domain of just size 1 already gives $c = 8$ (the first flipped spin costs energy $\Delta E = 6J$, and the second costs $\Delta E = 2J$). Since the $e^{c \b J}$ factor can compete with the $e^{O((\b J)^2)}$ factor at modest values of $\b J$, we will find it convenient to get rid of the former by considering dynamics where the nucleation of an initial minority domain is not rare. We do this by modifying the dynamics to instantly insert a minority domain of size 2 whenever the system reaches a state with maximal or minimal magnetization. We write $\wt t_{\sf mem}$ for the memory time under this dynamics, and plot $\wt t_{\sf mem}$ against $(\b J)^2$ in Fig.~\ref{fig:quantfig}~{\bf c}. The data are consistent with the $e^{O((\b J)^2)}$ scaling of~\eqref{main_result} as long as $\b J \gtrsim \sqrt 2$, and linear fits of $\ln \wt t_{\sf mem}$ to $(\b J)^a$ have lower $R^2$s for $a = 2$ than $a=1$ for all values of $v$ (see~\cite{code} for details).

	\paragraph*{Frictional cooling: } How should $E(v)$, the expected value of $H(t)$ in the nonequilibrium steady-state at shearing velocity $v$~\footnote{Which, incidentally, is not the Gibbs state of any local Hamiltonian~\cite{hilhorst2011two}.}, depend on $v$? Shearing minority domains apart temporarily leads to increases in energy, but if the sliding dynamics produces on net a smaller number of minority domains, it has a chance to nevertheless lead to a smaller value of $E(v)$. We show the result of measuring $E(v)$ numerically in Fig.~\ref{fig:quantfig}~{\bf d}, which shows that as long as $\b J$ and $v$ are large enough, $E(v)/E(0)$ decreases as a function of $v$. 
	Usually, frictional forces between two sliding systems heat these systems up, and this is indeed what happens in our system at small $\b J,v$. However, in the regime where sliding increases $\tmem$, the opposite is true: at least as judged by the steady-state energy, frictional forces instead cool the system down (see also the discussion in~\cite{kadau2008magnetic}). 
	
	\paragraph{Outlook: } 
	This work has shown that a very simple nonequilibrium drive---sliding two systems past one another at speed $v$---can dramatically slow down relaxation in a locally-interacting 1d spin system, increasing the memory time from $e^{O(\b J)}$ at $v=0$ to $e^{O(v(\b J)^2  )}$ at $v>v_c$, where $v_c \leq 1$. This increase happens because the critical size $\ler$ of a minority domain that is not eroded by the dynamics increases from $\ler = 1$ at $v=0$ to $\ler \sim v e^{\b J}$ at large enough $\b J, v$. 
	
	How slow can we go? Two things in this model limit how large $\tmem$ can get. The first is the scaling of $\ler$ with $\b J$: is it possible to design models where $\ler$ grows even faster? Additionally, in our model, the creation of a critical domain is suppressed in probability only by a factor of $e^{-O(\b J \ln \ler)}$: are there models where this suppression instead scales as $e^{-O(\b J \ler)}$? Answering either of these questions in the affirmative would move us even closer to constructing a simple model with long-range order in a locally-interacting 1d system.  
	Beyond this, it will also be interesting to examine other types of sliding in two and higher dimensions, where sliding is known to modify critical points~\cite{hucht2009nonequilibrium,angst2012strongly}, and to consider generalizations with different symmetry groups~\cite{igloi2011nonequilibrium}. Finally, it is natural to wonder what other nonequilibrium phenomena can be realized in this setting. Reference~\cite{stahl2026brownian} recently constructed simple nonequilibrium drives that can simulate {\it arbitrary} types of active dynamics, and it will be interesting to understand if similar results hold for sliding and related mechanisms. 
	
	\paragraph*{Acknowledgements: } 
	We thank Gesa D{\"u}nnweber, Yaodong Li, David Limmer, and Nathaniel Selub for discussions. C.S. was supported by the Office of Naval Research Young Investigator Program (ONR YIP) under Award Number N000142412098. E.L. was supported by a Miller Research Fellowship. All code supporting the numerical analysis in this work is available at~\cite{code}. 
	
	\bibliography{paper}
	
	\appendix 
	
	\
\end{document}


\title{Supplementary Material}
	\author{Ethan Lake}
	\date{\today} 
	\maketitle 

\begin{figure*}
	\centering 
	\includegraphics[width=.7\tw]{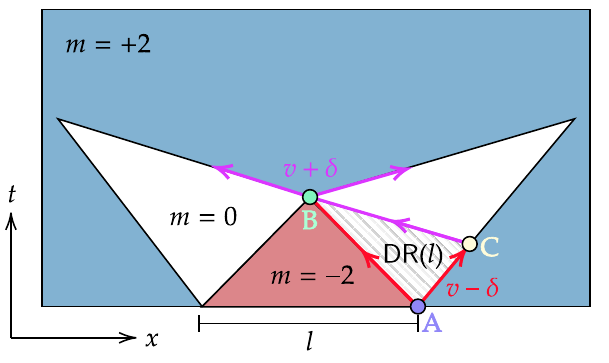}
	\caption{\label{fig:tlv_fig} How a minority domain of size $l$ is eroded under ideal sliding dynamics at $\b \ra \infty$, which maps to a version of Toom's two-line-voting automaton. The white regions have zero magnetization, and are metastable. The red arrows indicate interfaces that move with expected speed $v-\d$, and the purple arrows those that move with expected speed $v+\d$. The danger region $\dr(l)$ indicated by the triangle $\bfA\bfB\bfC$ is the spacetime region where spin flips in the metastable zero-magnetization area can propagate to prevent the erosion of the $m=-2$ (red) part of the initial minority domain.  } 
\end{figure*}

This file contains supplementary information supporting the main paper.

\tableofcontents

\section{Size of the danger region} \label{sec:danger}

In this section we derive a simple formula for the size of the danger region quoted in the main text. Consider the erosion of a minority domain of size $l$ under $\b \ra \infty$ sliding dynamics. The spacetime history of such a domain's erosion is shown in Fig.~\ref{fig:tlv_fig}. One of the danger regions $\dr(l)$ is shown as the triangle $\bfA\bfB\bfC$. 

Putting the origin of coordinates at the center of the minority domain, the spacetime coordinates of the points $A,B$ in Fig.~\ref{fig:tlv_fig} are, by simple geometric considerations, 
\bea (t_A,x_A) & = (0,l/2) \\ 
(t_B,x_B) & = \(\frac l{2(v-\d)},0\),\eea 
and the coordinates of $C$ are determined from the two equations 
\bea \frac{x_C - l/2}{v-\d} & = t_C  \\ 
\frac{x_C}{v+\d} & = t_B - t_C,\eea
which gives 
\be x_C-l/2  = \frac{l\d}{2v}.\ee 
Therefore the area of the danger region is 
\bea |\dr(l)| & = \frac12 |\bfA\bfC \times \bfA\bfB| \\ & = \frac12 | (x_C-l/2),t_C) \times (-l/2 , t_B)| \\ & = \frac{l^2 \d}{4v(v-\d)},\eea 
giving the result quoted in the main text. 

\section{More data on the scaling of $t_{\sf mem}$}

In this section, we provide more detail regarding what the numerics say about the scaling of the memory time with $\b J$ and $v$. 

First for the scaling with $\b J$. For the reasons explained in the main paper, we examine the nucleation-assisted memory time $\wt t_{\sf mem}$, with automatic injection of size 2 domains occurring in the maaximally magnetized states. 
Running a linear fit of $\ln \wt t_{\sf mem}$ to $(\b J)^2$ gives
\be \igpfc{tmemtilde_vs_r_a2fit} \ee 
where the bottom panel shows the residuals of the fit. On the other hand, running a linear fit of $\ln \wt t_{\sf mem}$ to $\b J$ gives 
\be \igpfc{tmemtilde_vs_r_a1fit} \ee 
with clearly worse residuals and a systematic curvature away from a linear fit. 

For the scaling with $v$, we re-plot the data in the main text, but now with the residual against a linear fit of $\ln \wt t_{\sf mem}$ to $v$. This gives 
\be \igpfc{tmem_vs_v_with_residual} \ee 
showing that the fit becomes good quality when $v > 2$, as expected.

\section{Phase diagram in the $v\ra \infty$ limit}

When $v \ra \infty$, the dynamics enters a simple mean-field limit: one chain couples randomly to sites of the other, seeing only (in expectation) the other chain's average magnetization (see also e.g.  \cite{kadau2008magnetic}). Since the two chains are statistically the same, we expect the dynamics in this limit to be governed by thermal dynamics with respect to the time-{\it independent} Hamiltonian 
\be H = - \sum_i \( (J \lan s\ran + h) s_i + J s_is_{i+1}\)\ee
where we have included an additional external magnetic field $h$ for completeness.  
We then solve for $\lan s\ran$ using a self-consistency equation derived from the exact expression for the average magnetization of a 1d Ising model in a field $\eta=J \lan s \ran + h$. This is easily arrived at using the transfer matrix approach. Recall that the transfer matrix is 
\be T = \bpm e^{\b (J+\eta)} & e^{-\b J} \\ e^{-\b J} & e^{\b(J-\eta)} \epm,\ee 
and the free energy density is 
\be F = -\frac1\b \ln \l_+,\ee 
with $\l_+$ the largest eigenvalue of $T$. Some algebra gives 
\be F = -J - \frac1\b \ln\( \cosh(\b \eta) + \sqrt{\cosh^2(\b\eta) - 2e^{-2\b J} \sin(2\b J)}\).\ee 
Taking the derivative with respect to $\eta$ then gives the magnetization 
\be m = \frac{\sinh(\b \eta)}{\sqrt{\cosh^2(\b \eta) - 2e^{-2\b J} \sinh(2\b J)}}.\ee 
In the present mean-field model, we thus have the self-consistent relation 
\be \lan s \ran = \frac{\sinh(\b (J \lan s\ran + h))}{\sqrt{\cosh^2(\b (J \lan s \ran + h) - 2e^{-2\b J} \sinh(2\b J)}}.\ee 
Working in units where $J=1$ then gives the following phase diagram: 
\be \igpfc{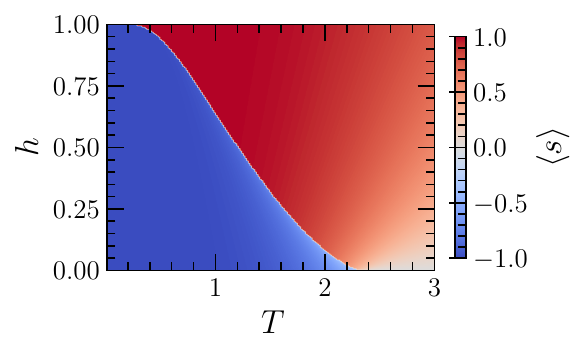}\ee 
where we show the value of $\lan s\ran$ in the less stable minimum (if multiple minima exist). 

\section{Details about forward-flux sampling}
\label{sec:ffs}

At low enough $T$ and large enough $v$, our prediction for the scaling of $\tmem$ rapidly becomes so large that confirming this prediction with a direct Monte-Carlo (MC) measurement of $\tmem$ is infeasible. We instead estimate $\tmem$ using forward-flux sampling (FFS) \cite{allen2009forward}, a rare-event method that factorizes the long crossing time into a product of conditional crossing probabilities, each of which is $O(1)$ and can be measured directly. This section spells out how FFS is implemented in our code and how the various tuning parameters are chosen. All  source files referenced below are in the public repository \cite{code} (the FFS routines themselves live in \texttt{src/ffs.jl}).

\ss{Setup}

We use the (signed) total magnetization per spin as the FFS order parameter (reaction coordinate),
\be \label{eq:m_def} m \equiv \frac{1}{2L} \sum_{i=1}^L \( s^{(\sft)}_i + s^{(\sfb)}_i \) \in [-1,+1].\ee
Our simulations use single-spin Glauber dynamics at inverse temperature $\b$, interleaved with the sliding shift: in our time units, one time step consists of $2L$ attempted spin flips interleaved with $v$ rigid translations of the top chain by one lattice site, with the translations spaced at regular intervals (see \texttt{src/core.jl}).

The metastable basin $A$ consists of configurations near the all-$+$ state, and the absorbing basin $B$ consists of configurations with $m \leq m_*$, where $m_*$ is the \texttt{ M\_threshold} parameter in the code, which we take to be $1/2$ in numerics. 

\medskip 

The intuition behind FFS is that even when the full barrier-crossing probability per unit time is very small in some control parameter, the {\it conditional} probability of advancing a small step up the barrier---given that the system has already made its way partway up---is of order unity. Concretely, we introduce a sequence of non-intersecting interfaces
\be \label{eq:interfaces} m_0 > m_1 > m_2 > \cdots > m_n = m_*\ee
in the $m$, lying between basin $A$ and basin $B$. The mean first-passage time $\tmem$ from $A$ to $B$ then admits the standard FFS factorization \cite{allen2009forward}
\be \label{eq:ffs_master} \tmem^{-1} \approx \Phi_0 \prod_{k=1}^n P_k,\ee
where $\Phi_0$ is the flux of forward crossings of $m_0$ in trajectories confined to basin $A$, and $P_k$ is the conditional probability that a trajectory which has just crossed $m_{k-1}$ for the first time subsequently reaches $m_k$ before returning to basin $A$. Each $P_k$ and $\Phi_0$ is by construction $O(1)$, so the right-hand side of \eqref{eq:ffs_master} can be quickly evaluated even when $\tmem$ itself is very large.


\ss{Phase 0: initial flux}

The first ingredient is the basin-$A$ flux $\Phi_0$ through the first interface $m_0$. We initialize a configuration in the all-$+$ state and let it equilibrate within basin $A$ for a short time (see \texttt{measure\_initial\_flux} in \texttt{src/ffs.jl}). We then run a long trajectory, monitoring $m(t)$ at every integer time step. Each time $m$ crosses $m_0$ from above we record the configuration (this is a ``forward crossing''); after such a crossing we wait for $m$ to return above $m_0$ before counting the next one, so that consecutive crossings are independent excursions. If during one of these excursions $m$ falls below the absorbing threshold $m_*$ we terminate the trajectory and start a new one in basin $A$. Calling $\mathcal{T}$ the total time spent in basin $A$ across all such trajectories and $N_0$ the number of forward crossings, we estimate
\be \label{eq:phi0} \Phi_0 \approx N_0 / \mathcal{T}, \qquad \mathrm{Var}(\ln \Phi_0) \approx 1/N_0.\ee
The ensemble $\mathcal{C}_0 = \{C_0^{(1)}, \dots, C_0^{(N_0)}\}$ of recorded configurations serves as the source for the next phase. We run our simulations until the size of each $\mcc_k$ reaches a fixed value $n_{\rm configs}$, set with the \texttt{--n\_configs\_per\_run} command-line argument in the code. We use $n_{\rm configs} \sim 15,000$ for the numerical results in the paper. 

\ss{Placement of the first interface}

The first interface $m_0$ should be far enough into basin $A$ that crossings are reasonably frequent, but far enough from the all-$+$ state that the recorded configurations are genuinely partway up the barrier rather than typical fluctuations within $A$. We choose the interface magnetization adaptively by first measuring the metastable mean magnetization $\overline{m}_A$ and its standard deviation $\sigma_A$ from a brief (ten time steps) run started in the all-$+$ state, and then setting
\be m_0 = \overline{m}_A - 2.5 \, \sigma_A.\ee
This puts $m_0$ in the lower tail of the metastable distribution: crossings occur often enough that $\Phi_0$ converges quickly, but the configurations recorded already represent fluctuations distinguishable from typical $A$-states (if $m_0 \le m_*$ then the transition is not actually a rare event and the FFS estimate is reported as failed).

\ss{Phase $k$: interface advancement}

Given a source ensemble $\mathcal{C}_{k-1}$ at interface $m_{k-1}$, we measure $P_k$ by firing trial trajectories from configurations drawn uniformly from $\mathcal{C}_{k-1}$. Each trial evolves under the same dynamics, with two stopping conditions:
\begin{itemize}
\item {\bf success}: the trial reaches $m \le m_k$, and its endpoint configuration is appended to the new source ensemble $\mathcal{C}_k$.
\item {\bf failure}: the trial returns above a {\it failure boundary} $m_{\rm fail}$, defined below.
\end{itemize}
We continue firing trials until $|\mathcal{C}_k| = n_{\rm configs}$ successful endpoint configurations have been collected. Letting $N_k^{\rm tot}$ be the total number of trials fired at phase $k$ (successes plus failures),
\be P_k \approx n_{\rm configs}/N_k^{\rm tot}, \quad \mathrm{Var}(\ln P_k) \approx (1-P_k)/(P_k N_k^{\rm tot}).\ee

It occasionally proves numerically helpful to use a {\it lookback failure boundary} when defining $m_{\rm fail}$: rather than declaring failure whenever a trial returns above $m_0$, we declare failure only when the trial returns above $m_{\max(k-K,0)}$, where $K = 7$ is a fixed integer (\texttt{n\_lookback} in the code). 

\ss{Adaptive interface placement}

Equation \eqref{eq:ffs_master} is exact for any choice of interfaces, but the variance of the estimator depends strongly on their placement. The optimal choice is to equalize the $P_k$s, since for fixed total computational budget the relative error is minimized when each interface costs the same number of trials. We achieve this by placing each new interface adaptively.

Given the source ensemble $\mathcal{C}_{k-1}$ at interface $m_{k-1}$, we fire $n_{\rm configs}$ exploratory trials. Each trial runs until it either falls below $m_*$ or returns above the lookback failure boundary, and we record the minimum magnetization $m_{\min}^{(j)}$ reached by trial $j$. Sorting $\{m_{\min}^{(j)}\}$ in ascending order and letting $\alpha \in (0,1)$ be a target conditional crossing probability (the \texttt{--target\_crossing\_prob} command-line argument; we use $\alpha = 0.25$ by default), we set
\be m_k = m_{\min}^{(\lceil \alpha n_{\rm configs} \rceil)}.\ee
By construction, a fraction $\approx \alpha$ of trajectories launched from $\mathcal{C}_{k-1}$ will reach $m_k$ before returning, so $P_k \approx \alpha$ to within sampling noise. 

As an alternative, the code also supports a {\it canonical} mode (the \texttt{--n\_interfaces} command-line argument set to a positive integer) in which the interfaces are placed uniformly between $m_0$ and $m_*$ and $m_{\rm fail} = m_0$ is held throughout; this is useful in regimes where there is no genuine metastable barrier, but is not used for any data shown in the main text.

\ss{Combining the phases}

Once all phases have run, the estimator \eqref{eq:ffs_master} gives
\be \label{eq:tmem_estimator} \log \tmem = -\log \Phi_0 - \sum_{k=1}^n \log P_k\ee
with statistical variance
\be \label{eq:tmem_var} \mathrm{Var}(\ln \tmem) \approx \frac{1}{N_0} + \sum_{k=1}^n \frac{1-P_k}{P_k N_k^{\rm tot}}.\ee
The dominant cost of an FFS estimation is the number of interfaces $n$ times the number of trials per interface. For the data shown in the main text we use $n_{\rm configs} = 15,000$ and usually have around 10-20 interfaces; with these choices, the data displayed in the main text can be obtained in a few tens of minutes on a laptop. 

\ss{Isolating the nucleation rate} \label{ss:isolating_nucleation_rate} 

As discussed in the main text, relaxation in the systems we are concerned with is nucleation-limited rather than aggregation-limited. 
The logic is as follows: once a supercritical droplet forms at size $\ler$, it grows after a time $t$ to size $\sim \ler + \sqrt{D t}$, coditioned on its size not dropping below $\ler$ at any point; here $D$ is the (noise-dependent) diffusion constant of the random walk executed by a well-isolated domain wall. The probability that the size of the droplet does not drop below $\ler$ by time $t$ is however $\sim 1/\sqrt{D t}$, so the total size the droplet contributes in expectation is independent of $D$ and $t$. 

This means that the scaling of $\tmem$ is set by the nucleation time of a supercritical droplet, and is independent of $D$ to leading order. 
To estimate the scaling of $\tmem$ most efficiently in FFS numerics, we therefore compute the memory time on a system of size $L = c \ler$, where $c$ is a constant of order unity (we fix $c = 5$ in our numerics). 

\ss{Independent repetitions and error bars}

To check that the variance estimate \eqref{eq:tmem_var} is not dominated by tail effects or by an unlucky choice of source ensembles, we repeat the entire FFS procedure $n_{\rm rep}$ times for each choice of parameters point (this is the \texttt{--n\_repeats} command-line argument; we use $n_{\rm rep} =5$ for the data shown in the main text), each repeat using independent seeds and independent adaptive interface placements. We report the mean of $\log_{10} \tmem$ across the $n_{\rm rep}$ runs, with error bars given by
\be \sigma_{\log \tmem} = \max \( \frac{1}{\ln 10}\sqrt{ \frac{\overline{\mathrm{Var}(\ln \tmem)}}{n_{\rm rep}} }, \, \frac{\mathrm{std}(\log_{10} \tmem)}{\sqrt{n_{\rm rep}}} \) ,\ee
where the first argument is the (square root of the) mean intrinsic FFS variance \eqref{eq:tmem_var}  and the second is the empirical standard error across the $n_{\rm rep}$ repeats. 

\section{Memory in the GKL automaton}

\ss{Setup and background}

In this section we investigate the relaxation dynamics of the noisy GKL automaton \cite{gacs1978one} under noise of strength $\ep \in [0,1]$ and bias $\eta \in [-1,1]$. In this dynamics, the spins all update synchronously in discrete time, and each spin is subjected to an error with probability $\ep$. When an error occurs, a spin is replaced by $\pm1$ with probaiblity $(1\pm \eta)/2$.

It was proven in \cite{park1997ergodicity} that there is a positive constant $\eta_c>0$ such that 
\be \tmem = e^{\ct(\log^2 \ep)} \qq \forall \, \,\eta \in [-1, -\eta_c] \cup [\eta_c , 1],\ee 
where $\ct$ denotes scaling with respect to $\ep$. The general proof strategy is quite simple to summarize: following a similar argument as in Sec.~\ref{sec:danger}, one concludes that the erosion length in the GKL automaton is, in the $\ep\ll1$ limit 
\be \label{gkller} \ler \sim \frac1{\sqrt \ep},\ee 
where the $\sim$ omits $\ep$-independent factors. 
Since an error that occurs in the danger region can grow a minority domain of size $l$ to one of size $\l l$ with $\l > 1$, growing a domain to size $\ler$ only requires $\log_\l \ler$ steps. Thus 
\be \label{gkl_tmem} \tmem \sim \ep^{-\log_\l \ler} = e^{-\ct(\log ^2\ep)}.\ee 

While Ref.~\cite{park1997ergodicity} did not make rigorous statements about the case of unbiased noise, the physical arguments above do not depend crucially on taking $\eta \neq 0$. However, in \cite{de1992gacs} it was proposed from direct Monte Carlo numerics that in fact $\tmem = e^{\ct(1/\ep)}$ when $\eta = 0$. Unfortunately, distinguishing between $e^{\ct(1/\ep)}$ and $e^{\ct( \log^2 \ep)}$ is very difficult in naive Monte Carlo, especially with the limited range of $\ep$ values accessed in \cite{de1992gacs}. In what follows, we revisit this question with more precise numerics. 

\ss{Scaling of the erosion length} 

First, we numerically confirm that \eqref{gkller} holds regardless of $\eta$. This is done using the same procedure as used for the sliding Ising model, and gives 
\be \igpfc{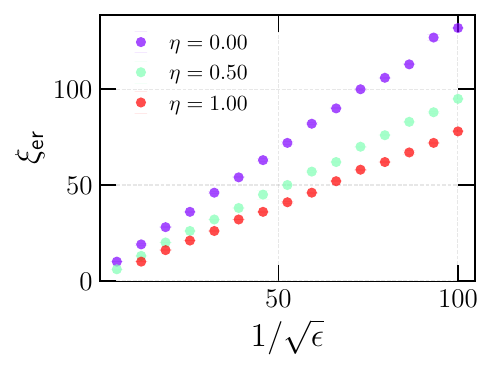} \ee 
as predicted. 

\ss{Domain wall diffusion} 

Once formed, a supercritical droplet of size $l > \ler$ has endpoints which undergo biased diffusion, with the sign of the bias determined by $\eta$. We can numerically estimate $D(\ep)$ by preparing a system in a state with $s_{i<L/2} = +1, s_{i\geq L/2} = -1$, and computing the diffusion constant of the domain wall position $x_{dw} = \sum_i (s_i + 1)/2$. Doing this at zero bias and with a system of size $L=4000$ gives $D(\ep) \propto 1/\sqrt\ep$:  
\be \igpfc{gkl_diffusion_constant}.\ee 
The behavior of $D$ is tangential to the discussion of $\tmem$ however, since relaxation is nucleation-limited rather than coalesence-limited, for the same reason as in the sliding Ising chain context, c.f. Sec.~\ref{ss:isolating_nucleation_rate}. 

\ss{Scaling of $\tmem$} 

We now numerically check \eqref{gkl_tmem} with the FFS technique, using the adaptive system size protocol of Sec.~\ref{ss:isolating_nucleation_rate}. We set $m_*=1/2$ and $n_{\rm configs} = 15,000$. 
Plotting $\tmem$ against $(\log(1/\ep))^2$, we find 
\be \igpfc{gkl_tmem_vslog} \ee 
The same data plotted against $1/\ep$ clearly reveal that \eqref{gkl_tmem} fits the data better: 
\be \igpfc{gkl_tmem_vsq} \ee 

\bibliographystyle{apsrev4-2}
\bibliography{paper}

%% file: ethans_preamble.tex
\usepackage{graphicx}
\usepackage[nointegrals]{wasysym} %
\usepackage[export]{adjustbox}
\usepackage{amsmath,amsfonts,amssymb,latexsym}
\usepackage{hhline}
\usepackage{bm}
\usepackage{verbatim}
\usepackage{enumitem}
\usepackage{mathrsfs}
\usepackage{slashed}
\usepackage{empheq}

\newcommand{\ra}{\rightarrow}

\newcommand{\wt}{\widetilde}

\renewcommand{\[}{\left[}

\newcommand{\mt}{\mapsto}

\newcommand\bpm            {\begin{pmatrix}}
	\newcommand\epm           {\end{pmatrix}}

\def\app#1#2{%
	\mathrel{%
		\setbox0=\hbox{$#1\sim$}%
		\setbox2=\hbox{%
			\rlap{\hbox{$#1\propto$}}%
			\lower1.1\ht0\box0%
		}%
		\raise0.25\ht2\box2%
	}%
}

\newcommand{\tw}{\textwidth}

\newcommand{\ct}{\Theta}

\newcommand{\inv}{^{-1}}

\newcommand{\ope}\odot

\usepackage{manfnt}

\newcommand{\bi}{\begin{itemize}}
	\newcommand{\ei}{\end{itemize}}

\usepackage{amsthm}

\newtheorem{proposition}{Proposition}

\theoremstyle{definition}

\newcommand\bpro		  {\begin{proposition}}
	\newcommand\epro 		  {\end{proposition}}
\newcommand\bproof			  {\begin{proof}}
	\newcommand\eproof 		  {\end{proof}}

\newcommand\ed            {\end{definition}}

\newcommand\be            {\begin{equation}}
\newcommand\ee            {\end{equation}}
\newcommand\ba            {\begin{aligned}}
\newcommand\ea            {\end{aligned}}
\newcommand\bea{\begin{equation}\begin{aligned}}
	\newcommand\eea{\end{aligned}\end{equation}}

\usepackage{hyperref} 
\hypersetup{final}
\hypersetup{colorlinks, citecolor=red, linkcolor=red, urlcolor=red}

\renewcommand{\b}{\beta}
\renewcommand{\d}{\delta}

\newcommand{\g}{\gamma}

\newcommand{\s}{\sigma}

\newcommand{\ep}{\varepsilon} %

\renewcommand{\O}{\Omega}

\newcommand{\bfA}{\mathbf{A}}

\newcommand{\bfC}{\mathbf{C}}

\newcommand{\zt}{\mathbb{Z}_2}

\newcommand{\sfD}{\mathsf{D}}

\newcommand{\sfP}{\mathsf{P}}

\newcommand{\sfR}{\mathsf{R}}

\newcommand{\sfb}{\mathsf{b}}

\newcommand{\sft}{\mathsf{t}}

\usepackage[mathscr]{eucal} %

\usepackage{dcolumn}

\usepackage{pifont}
\usepackage[normalem]{ulem} 